\documentclass[letterpaper]{jpconf}
\usepackage{graphicx}
\begin{document}
\title{Spatial distribution of electrons on a superfluid helium charge-coupled device}
\author{Maika Takita$^1$, F R Bradbury$^{1{\dagger}}$, T M Gurrieri$^2$, K J Wilkel$^2$, Kevin Eng$^{2{\ddagger}}$, M S Carroll$^2$, and S A Lyon$^1$}

\address{$^1$ Department of Electrical Engineering, Princeton University, Princeton, NJ}
\address{$^2$ Sandia National Laboratories, Albuquerque, NM}
\address{$^{\dagger}$Present address: Amsterdam University College, Amsterdam, The Netherlands}
\address{$^{\ddagger}$Present address: HRL Laboratories, LLC, Malibu, CA}

\ead{mtakita@princeton.edu}

\begin{abstract}

Electrons floating on the surface of superfluid helium have been suggested as promising mobile spin qubits. Three micron wide channels fabricated with standard silicon processing are filled with superfluid helium by capillary action. Photoemitted electrons are held by voltages applied to underlying gates. The gates are connected as a 3-phase charge-coupled device (CCD). Starting with approximately one electron per channel, no detectable transfer errors occur while clocking $10^9$ pixels. One channel with its associated gates is perpendicular to the other 120, providing a CCD which can transfer electrons between the others. This perpendicular channel has not only shown efficient electron transport but also serves as a way to measure the uniformity of the electron occupancy in the 120 parallel channels. 
\end{abstract}

\section{Introduction}

Electrons are bound to the surface of superfluid helium by a weak image potential in addition to potentials from underlying electrostatic gates which control the position of these electrons on helium [1-4]. Electrons typically float in vacuum about 10nm above the surface, forming a clean two dimensional electron system (2DES). It is a well studied system which has shown the highest mobility of any 2DES [5], and exhibited the first 2D Wigner crystal [6]. 

Electron spins have been studied as possible quantum bits (qubits), since they are natural two level quantum systems, but their coherence is rapidly degraded during transport in semiconductors. Unlike other electron spins in heterostructure 2DES systems, spins on the surface of helium reside in vacuum and are expected to suffer much less decoherence, especially from spin-orbit interactions [7]. Hence, the possibility of mobile qubits in a condensed matter system arises. Using existing fabrication technology we have shown that manipulating a large number of spins with a small number of gates is possible [8].

\section{Experimental setup}

Devices with electrons confined in channels have been studied previously [9-13].  Similar to those, a standard complementary metal-oxide-silicon (CMOS) process at Sandia National Laboratory was used to fabricate 2$\mu$m deep, 3$\mu$m wide channels [14]. The upper metal interconnect layers are used to fabricate the channels and gates. The topmost metal layer defines the channel width, and acts as a mask for etching the surrounding dielectric layers (SiO$_2$) down to the underlying gate electrodes, running under 120 parallel channels (see Fig.1). Electrons are photoemitted onto the sample from a zinc film placed a few millimeters above the device. By keeping the underlying gates positively biased with respect to the zinc film, electrons are attracted to the channels [15]. The large gate labeled ``right reservoir" is where the electrons are initially collected. Electrons collected over the reservoir are blocked from moving above the small gates by keeping the ``door" gate dc-voltage negative (repulsive) with respect to the reservoir. Lowering the door gate potential allows a packet of electrons to moves from the reservoir to the electron detection gates, ``twiddle" and ``sense". These electrons are detected through their capacitive coupling to the sense gate, which is connected to the gate of a nearby high electron mobility transistor (HEMT). An ac-voltage is applied to the twiddle gate to push the electrons over the sense gate and pull them away. The sense gate sees a periodic charge induced voltage, which is buffered by the HEMT and detected using a lockin amplifier.

The small gates in the horizontal channels have a 3$\mu$m period including a 0.5$\mu$m gap. These underlying gates to the left of detection gates are connected as a 3-phase CCD, with every third gate connected by on-chip vias to its clock rail in a lower metal layer. Each set of three gates; $\phi1$, $\phi2$, and $\phi3$; makes up one pixel for electron control. A 2.5 $\mu$m vertical channel connecting 120 horizontal channels has two extra gate electrodes, Vccd$_2$ and Vccd$_3$. These two gates and $\phi$2 from the horizontal channel comprise a 3-phase CCD to bring electrons from one channel to the others. The device is wired and mounted in a cell that is cooled to 1.6K. The cell is filled with superfluid helium up to about 1cm below the device. Helium fills the channels by capillary action with the depth in the channel being determined by the device structure.

\begin{figure}[h]
\includegraphics[width=14pc]{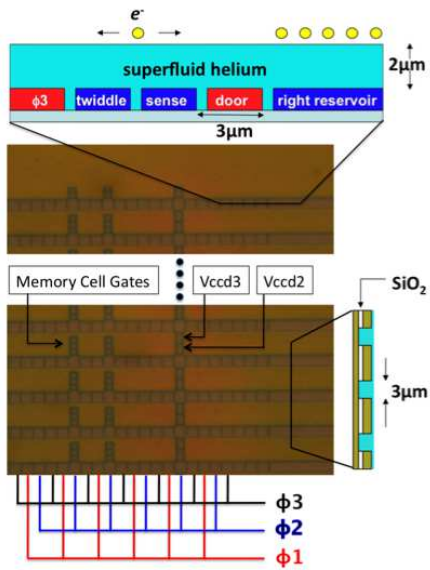}\hspace{2pc}%
\begin{minipage}[b]{22pc}\caption{\label{label}Seven of the 120 parallel horizontal channels are seen, with a schematic cross-section to the right of the image.  The darker regions of the image are the channels with underlying gates, and the lighter areas are the metal layer between the channels. Electrons are confined above the helium in the channels by a positive dc bias (3V) on the gates with respect to the top metal plane. The 3-phase CCD connection is drawn schematically below the micrographs. A cartoon above the channel image is a cut-away along the middle of a channel giving a scale view of the loading and detection gates. The horizontal channels are interposed by two memory cells for electron storage and an interconnecting vertical CCD channel. All gates used in these experiments run under and control electrons in all 120 channels in parallel.}
\end{minipage}
\end{figure}

\section{Electron transfer efficiency}

Unprecedented electron transfer efficiency has been demonstrated by clocking a packet of electrons back and forth along the horizontal channels for more than one billion pixels [8]. When loading electrons from the reservoir onto the small gates, we initially clear out the CCD pixels and make sure that electrons in the right reservoir are blocked with the door gate closed (negative). Electrons are loaded from the reservoir onto the twiddle and sense gates by opening and then closing the door. To achieve a desired pixel occupancy, the potential of the door gate is lowered temporarily to bleed off excess electrons back to the reservoir. After the initial charge measurement, a packet of electrons is moved into one pixel of the CCDs and clocked along the channels using an appropriate voltage sequence, similar to the 3-phase CCD sequence first described by Boyle and Smith[16]. The reliability of 2D transport in the channel array has also been tested using a vertical CCD channel. A packet of electrons in lower half of the channels (60 channels) are moved in a ``C" pattern: one pixel left, 60 pixels up, one pixel right, and then the sequence is reversed to return the electron packets to their starting point. No measurable failure has been seen after a billion pixel cycles.

\section{Channel uniformity}

The sense gate runs under all 120 horizontal channels and thus measures the sum of the electrons in all of them. The signal calibration in units of electrons per channel assumes perfect uniformity over 120 parallel channels. When we see a signal corresponding to 120 electrons, we cannot know whether there is one electron per channel or several electrons in some of the channels and none in the others. The vertical channel can be utilized in a measurement of the distribution of electrons amongst the horizontal channels. There are no reservoirs at the top or bottom of the array (note that up and down are defined here by the orientation of Fig. 1). By progressively shifting the electrons up or down, electrons are ejected. When the remaining electrons are moved across to the sense gate, the signal is reduced. 

As long as the devices are not overcharged during the initial photoemission, bringing the underlying gates to the potential of the metal top plane serves to kick electrons out of the channels, probably out onto the thin van der Waals helium film covering the top metal plane. It is likely that these ejected electrons are immobilized by the roughness of the metal in this area, though for the analysis of these data it is sufficient to know, experimentally, that the ejected electrons do not return to the channels. 

Figure 2(a) plots the results of two vertical clocking experiments where electrons are clocked up or down 120 pixels. One pixel cycle using $\phi$2, Vccd$_2$, and Vccd$_3$ brings a packet of electrons up or down one channel. In the experiment, an electron packet in each channel is loaded from the reservoir, measured over the detection gates, and then clocked in the horizontal channels to the vertical CCD, over $\phi$2. This packet of electrons is clocked 12 pixels up or down and then brought back horizontally to the detection gates for a new measurement. This cycle continues until electrons from all 120 horizontal channels have been ejected at the top or bottom of the device. The data shows that all the groups of 12 channels, except the topmost 12, have approximately equal populations. The experiments were done after a specific loading (by photoemission) of the right reservoir. By repeating such an experiment for many reservoir loading levels, we find that the population of all 120 channels are equalized when the right reservoir has been well-filled by photoemission. 


\begin{figure}[h]
\includegraphics[width=14pc]{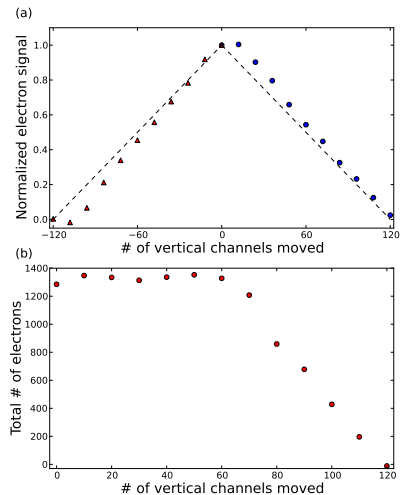}\hspace{2pc}%
\begin{minipage}[b]{22pc}\caption{\label{label}(a) Measurement of the electron population of the horizontal channels for a typical initial loading of electrons.  The circle (blue) and triangle (red) data points represent electron signals as the vertical channel is clocked upwards and downwards, respectively.  The electron signal is normalized to the initial loading; about 1600 electrons, or approximately 13.5 electrons per channel.  The dashed black lines show a linear depopulation which would correspond to perfctly uniform channel occupation. (b) Total number of electrons in all the channels after the C-pattern experiment. The electron signal is measured after each time the electrons were clocked up 10 pixels. A constant signal through the first 60 pixel cycles and then a linear decay confirm a uniform occupancy of the lower 60 channels.}
\end{minipage}
\end{figure}

After the C-pattern transport efficiency experiment described above, electrons occupying the lower 60 channels were brought to the vertical channel to measure the charge uniformity. Electrons are clocked 10 pixels up and then brought across to the detection gates. The electron signal does not decrease for the first 60 pixel cycle which says that the top 60 channels are unoccupied, as expected (these channels were emptied by the C-pattern). The electron signal decreases linearly after 60 pixel cycles, which confirms the uniformity of bottom 60 channels (see Fig. 2(b)).

\section{Conclusion}
Manipulating a large number of qubits is necessary for applications in quantum computing. The charge-coupled device with electrons on the surface of helium has shown undetectably rare electron transfer failure, allowing electrons to be positioned at any of more than four thousand distinct sites by controlling only five electrostatic gates. This interpretation rests on the assumption that all the channels are approximately equally populated. Non uniformity could affect the interpretation of the clocking result, and it raises the question as to whether we can expect uniform occupancy for QC applications. We have shown that channel uniformity is achievable with the electron loading methods we have established, and have verified it using the vertical channel.

\ack
	We would like to thank S. Shankar and G. Sabouret for helpful discussions and J. Donnal for the electronics to apply clock voltages to the CCD. Work at Princeton was supported by the NSF under Grant Nos. CCF-0726490 and DMR-1005476. Additionally, this work was performed, in part, at the Center for Integrated Nanotechnologies, a U.S. Department of Energy, Office of Basic Energy Sciences user facility. Sandia is a multiprogram laboratory operated by Sandia Corporation, a Lockheed Martin Co., for the United States Department of Energy under Contract No. DE-AC04-94AL85000. Conference travel was supported by the NSF under Grant No. DMR-0844115 and ICAM-I2CAM, 1 Shields Avenue, Davis, CA 95616.

\end{document}